\documentclass{article}
\usepackage[T1]{fontenc}
\usepackage[utf8]{inputenc}
\usepackage{ismir} 
\usepackage{amsmath,cite,url}
\usepackage{graphicx}
\usepackage{color}
\usepackage{placeins}

\usepackage{booktabs}
\usepackage{graphicx}
\usepackage{array}
\title{Understanding Automatic Mixing: A Subtask-Oriented Analysis of Two-Stage Mixing System}

\multauthor
  {Jinjie Shi$^1$ \hspace{0.6cm}
   Wei Hua$^2$ \hspace{0.6cm}
   Kunzhu Xie$^3$}
  {{\bf
   Make Li$^4$ \hspace{0.6cm}
   Yuchen Liu$^5$ \hspace{0.6cm}
   Joshua Reiss$^1$}\\
   $^1$ Queen Mary University of London, United Kingdom\\
   $^2$ Guangxi Arts University, China\\
   $^3$ Wuhan University of Communications, China\\
   $^4$ Xinghai Conservatory of Music, China\\
   $^5$ University College London, United Kingdom\\
   {\tt\small jinjie.shi@qmul.ac.uk}
  }

\def\authorname{J. Shi, W. Hua, K. Xie, M. Li, Y. Liu, and J. Reiss}
\usepackage[bookmarks=false,pdfauthor={\authorname},pdfsubject={\pdfsubject},hidelinks]{hyperref}

\begin{document}

\maketitle

\begin{abstract}
Automatic mixing transforms multitrack recordings into perceptually coherent, balanced, and aesthetically consistent mixes. In real-world production, this task is challenging due to large track counts, diverse instrumentation, and strong inter-track dependencies. Two-stage systems address this complexity by separating intra-group processing from inter-group mixing, yet it remains unclear whether their gains arise from stronger component models or from explicit task decomposition. We present a subtask-oriented analysis of automatic mixing through three controlled listening experiments. We investigate whether full-mix models transfer to intra-group mixing, whether downstream models compensate for grouping and loudness errors, and whether two-stage decomposition improves full-mix quality. Across three dense pop and rock excerpts, transfer differs between the evaluated models; inappropriate grouping causes clear downstream degradation, while altered loudness relationships have weaker and model-dependent effects. Both two-stage variants significantly outperform their corresponding single-stage baselines. These findings support explicit separation of local balance and global mix coordination as a useful design principle for automatic mixing. Code and audio examples are available online\footnote{
\url{https://sparrowreivun.github.io/TwoStageMixingAnalysis/}}.

\end{abstract}

\section{Introduction}\label{sec:introduction}

\begin{figure*}[t]
\centering
\includegraphics[alt={Two-stage automatic mixing framework showing grouping, intra-group processing, and inter-group mixing stages},width=\textwidth]{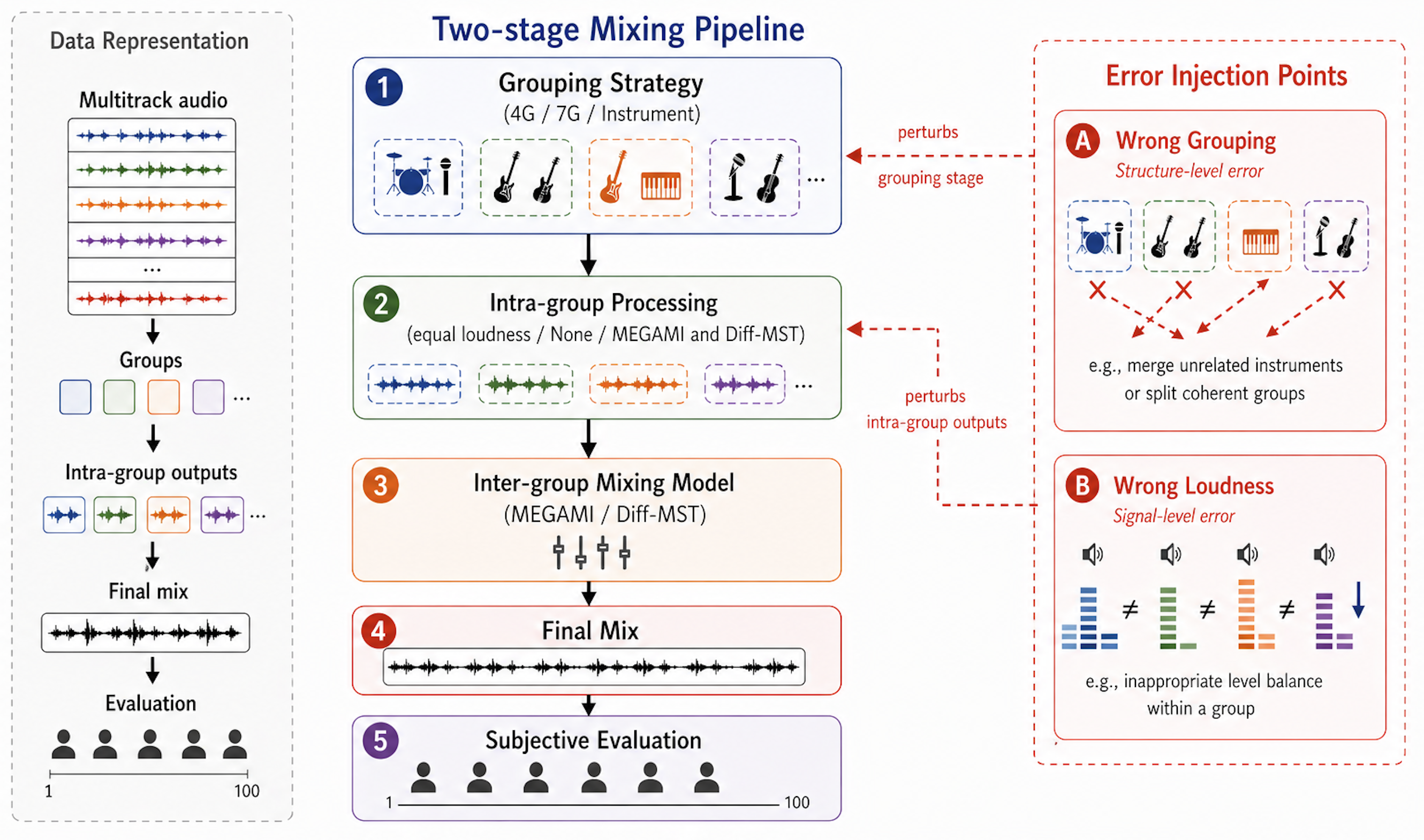}
\caption{Two-stage analysis framework. Controlled changes to grouping and intra-group processing are used to examine error propagation and downstream compensation.}
\label{fig:framework}
\end{figure*}

Automatic mixing must jointly control local balance, spectral and dynamic interactions, spatial organization, and overall style \cite{deman2019intelligent,wilmering2020history,moffatApproachesIntelligentMusic2019}. These relationships become increasingly difficult to model in productions containing many tracks and diverse instruments.

Recent data-driven methods commonly formulate mixing as a monolithic full-mix task. Some predict explicit effect parameters through differentiable mixing consoles \cite{vanka2024diffmst,steinmetz2020differentiable}, whereas generative systems model a distribution of plausible mixes for the same input \cite{moliner2025megami}.

Structured approaches instead divide mixing into grouping, local processing, and global coordination \cite{fenton2018automatic,wichernComparisonLoudnessFeatures2015,ronanImpactSubgroupingPractices2015,ronanAutomaticSubgroupingMultitrack2015,ronanAnalysisSubgroupingPractices2017}. Subgrouping can improve perceived clarity and quality \cite{ronan2018subgroups,shi2025scalable}, and hierarchical systems with separately trained subgroup models have demonstrated the feasibility of stage-specific processing \cite{koszewski2023automatic}. However, the contribution of decomposition itself remains insufficiently understood.

We compare three representative approaches. Equal local loudness (ELL) is a training-free, rule-based method used for intra-group balancing \cite{shi2025scalable}. Diff-MST predicts gain, equalization, compression, and panning parameters using a differentiable mixing console \cite{vanka2024diffmst}. MEGAMI uses conditional generative modeling to produce coordinated track-level effect representations \cite{moliner2025megami}. We use its released checkpoint, trained on examples containing up to 14 tracks; the Transformer architecture itself is not inherently restricted to this number. NoMix applies only the preprocessing shared by all conditions and serves as an unprocessed control.

We use the framework in Fig.~\ref{fig:framework} as an analysis scaffold rather than propose a new mixing system. We ask:

\textbf{RQ1:} Can models trained for full mixing transfer to intra-group mixing?

\textbf{RQ2:} Can downstream models compensate for incorrect grouping and loudness relationships?

\textbf{RQ3:} Does explicit two-stage decomposition improve full-mix quality?

The experiments reveal model-dependent transfer, clear sensitivity to inappropriate grouping, weaker effects of altered loudness relationships, and significant improvements of both evaluated two-stage variants over their corresponding single-stage baselines.

\section{Subtask-Oriented Analysis Framework}

\subsection{Two-Stage Decomposition}

Let the multitrack input be $\mathcal{X}={x_1,\ldots,x_N}$. A grouping function $g$ partitions the tracks into functional groups $\mathcal{X}^{(k)}$. Intra-group processing produces group-level stems, which are then passed to an inter-group model:
\begin{equation}
\mathcal{S}={F_{\mathrm{intra}}(\mathcal{X}^{(1)}),\ldots,
F_{\mathrm{intra}}(\mathcal{X}^{(G)})}, \qquad
y=F_{\mathrm{inter}}(\mathcal{S}).
\end{equation}
A monolithic model instead generates the final mix directly:
\begin{equation}
y=F_{\mathrm{full}}(\mathcal{X}).
\end{equation}

The framework allows grouping strategies, intra-group processors, and inter-group models to be varied independently. This supports controlled analysis of transfer, error propagation, and downstream compensation. Particular instantiations are denoted as two-stage systems (2S).

\subsection{Grouping Strategies}

We consider three grouping strategies:

\textbf{(1) 4-group baseline.}
A widely used grouping scheme in prior work and multitrack datasets \cite{bittnerMedleydbMultitrackDataset2014}, consisting of Bass, Drums, Vocal, and Other.

\textbf{(2) 7-group mixing-oriented grouping.}
We design a seven-group scheme specifically for group-based processing in automatic mixing. Starting from the commonly used 4-group organization, the scheme introduces functional distinctions that are relevant to mixing while keeping the overall design as simple as possible.

The Bass category remains a dedicated Low-Frequency group. The Drums category is divided into Low--Mid Percussion, which includes kick drums, snare drums, toms, and related components, and High Percussion, which includes hi-hats, cymbals, overheads, and other high-frequency percussive components.

Tracks in the original Vocal and Other categories are reassigned according to their functional roles in the mix. Lead contains lead vocals, solo instruments, principal melodies, and other perceptually dominant foreground tracks. Accompaniment contains backing and harmony vocals, as well as instruments that provide harmonic support, rhythmic accompaniment, or arrangement-level support. Pad contains sustained background layers, spatial textures, and tracks characterized by weak attacks or limited rhythmic definition. The residual Other group contains small effects, transitional sounds, and exceptional tracks that do not satisfy the definitions of the preceding groups.

Table~\ref{tab:grouping_comparison} summarizes the relationship between the 4-group baseline and the mixing-oriented 7-group scheme.

\begin{table}[t]
\centering
\caption{Relationship between the 4-group baseline and the mixing-oriented 7-group scheme. Vocal and Other tracks are reassigned according to their functional roles in the mix.}
\label{tab:grouping_comparison}
\small
\begin{tabular}{p{2.0cm}p{5.3cm}}
\toprule
4-group & Corresponding 7-group assignment \\
\midrule
Bass  & Low-Frequency \\
Drums & Low--Mid Percussion; High Percussion \\
Vocal & Lead; Accompaniment \\
Other & Lead; Accompaniment; Pad; Other \\
\bottomrule
\end{tabular}
\end{table}

\textbf{(3) Instrument-based grouping.}
A label-driven grouping derived directly from track metadata \cite{moliner2025megami}, used as a control condition. Unlike the mixing-oriented 7-group scheme, this strategy groups tracks according to instrument names rather than their functional roles in the mix. Consequently, the instrument-based grouping used in our experiments typically produced more than seven groups.

\section{Experimental Setup}

\subsection{Data and Participants}

All stimuli, together with the reference mixes used by Diff-MST, were drawn from the Cambridge Multitrack Library \cite{senior2018mixing}\footnote{\url{https://cambridge-mt.com/ms-mtk.htm}}. We selected three densely arranged pop and rock excerpts containing approximately 23, 24, and 27 tracks, respectively. Each listening-test item was approximately 15 seconds long. Although Diff-MST was trained on material from the same library, the selected source songs were added only after its training corpus had been established. We also checked the available training-data lists for both Diff-MST and MEGAMI and confirmed that none of the selected songs overlapped with their training data.

A total of 26 participants took part, including music enthusiasts and listeners with mixing or music-production experience. Samples were rated independently on a continuous 1--100
mixing-quality scale, adapted from established subjective
audio-evaluation practice \cite{itu2015bs1534}. Stimulus order was randomized.

Nineteen participants completed all items. Hidden repeated trials were used to assess intra-rater reliability, with Pearson correlation $r>0.75$ as the primary criterion. Eighteen participants were retained, yielding 1170 completed ratings. Hidden repeats were used only for reliability assessment and were excluded from the primary inferential analyses.

\subsection{Experimental Design and Statistics}

\noindent\textbf{Experiment 1 (RQ1)} compares ELL, MEGAMI, Diff-MST, and NoMix on intra-group mixing.

\noindent\textbf{Experiment 2 (RQ2)} examines whether downstream models can compensate for errors introduced during intra-group processing. Experiment 2a evaluates grouping errors by comparing instrument-based grouping, the 4-group baseline \cite{bittnerMedleydbMultitrackDataset2014}, and the proposed 7-group scheme under both MEGAMI and Diff-MST. Experiment 2b isolates the effect of intra-group loudness structure under a fixed 7-group assignment. In the \textit{with-balance} condition, ELL is applied within each group before the tracks are combined into group-level stems. In the \textit{no-balance} condition, the intra-group balancing step is bypassed, while the grouping assignments and all other preprocessing remain unchanged. The resulting seven group-level stems are then processed by the same downstream MEGAMI or Diff-MST model using identical model settings. Together, these comparisons test whether downstream models are sensitive to errors in upstream grouping and level balance, or whether such errors can be compensated for at the inter-group stage.

\smallskip
\noindent\textbf{Experiment 3 (RQ3)} compares baseline models (MEGAMI and Diff-MST) with their two-stage variants (2S-MEGAMI and 2S-Diff-MST) on full-mix tasks involving complex arrangements with more than 20 input tracks, and additionally include Human references. In the two-stage variants, the inter-group model remains unchanged, while its input is replaced by group-level representations obtained through grouping and intra-group processing. Based on stable patterns observed in Experiments~1--2, we adopt representative design choices (7-group and ELL) rather than selecting the best-performing configuration.

For direct full-mix inference, Diff-MST can process the original multi-track inputs without additional grouping. Although the Transformer architecture used by MEGAMI is not inherently restricted to a fixed number of input tracks, its publicly released checkpoint was trained on examples containing up to 14 tracks. Therefore, following the original MEGAMI preprocessing strategy, we apply label-driven grouping derived from track metadata \cite{moliner2025megami} to aggregate the input tracks into a compatible representation before MEGAMI inference.

Paired $t$-tests are applied to within-item comparisons, followed by FDR-BH correction \cite{benjamini1995controlling}. We report adjusted $q$-values, paired Cohen's $d_z$, and the number of paired listener--item observations. Full pairwise results are provided in the supplementary material.
    
\section{Results}

\subsection{Experiment 1: Intra-group Mixing Quality}

One Diff-MST output was silent; the corresponding item was excluded from all-method paired analyses, leaving two valid items and $n=36$ paired observations.

\begin{figure}[t]
\centering
\includegraphics[alt={Box plot comparing intra-group mixing quality across different models},width=\linewidth]{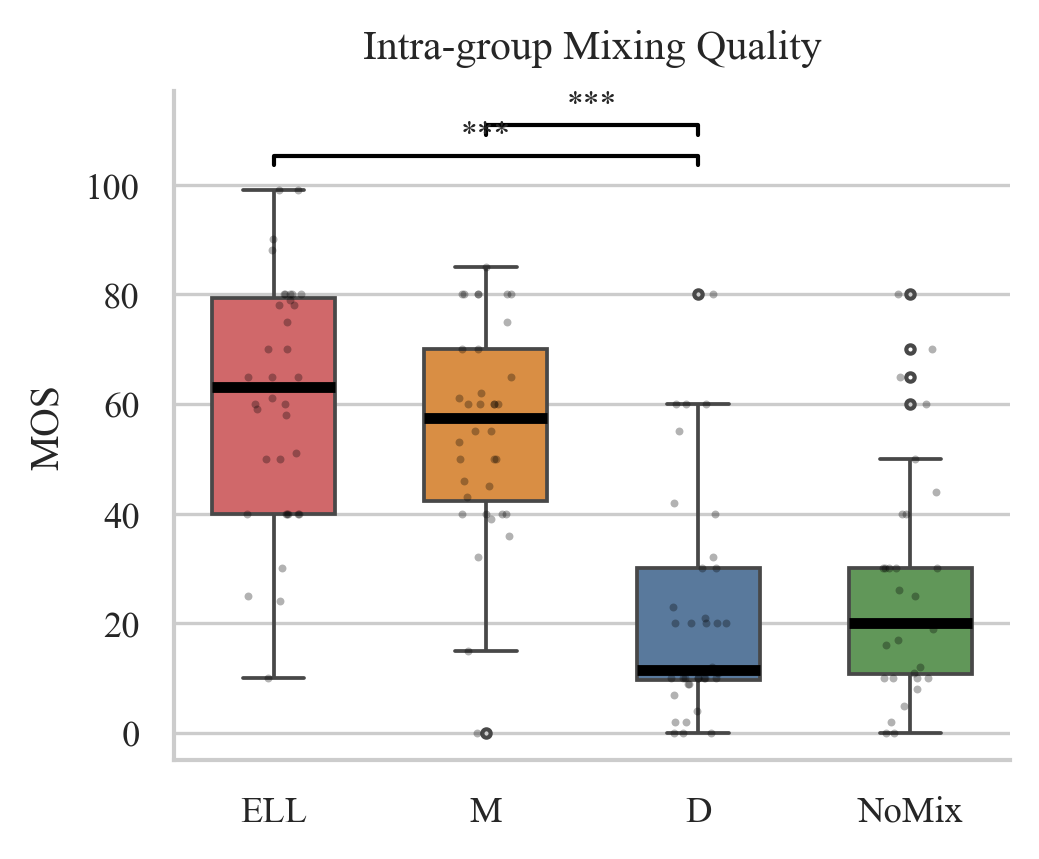}
\caption{Intra-group mixing quality (Experiment 1). ELL and MEGAMI outperform Diff-MST; full-mix models do not necessarily transfer equally to simpler subtasks. Significance markers indicate FDR-BH-adjusted values: * \(q<0.05\), ** \(q<0.01\), and *** \(q<0.001\).}
\label{fig:rq1}
\end{figure}

As shown in Fig.~\ref{fig:rq1}, ELL and MEGAMI receive higher ratings than Diff-MST and NoMix. ELL has the highest mean, but does not significantly outperform MEGAMI (61.08 vs.\ 55.47, $q=0.262$, $d_z=0.21$). MEGAMI significantly outperforms Diff-MST (55.47 vs.\ 21.36, $q=1.64\times10^{-9}$, $d_z=1.41$). Thus, transfer to the simpler intra-group task differs substantially between the evaluated full-mix models.

\begin{figure*}[t]
\centering
\includegraphics[alt={Box plots showing the effect of grouping and loudness errors on downstream compensation},width=\textwidth]{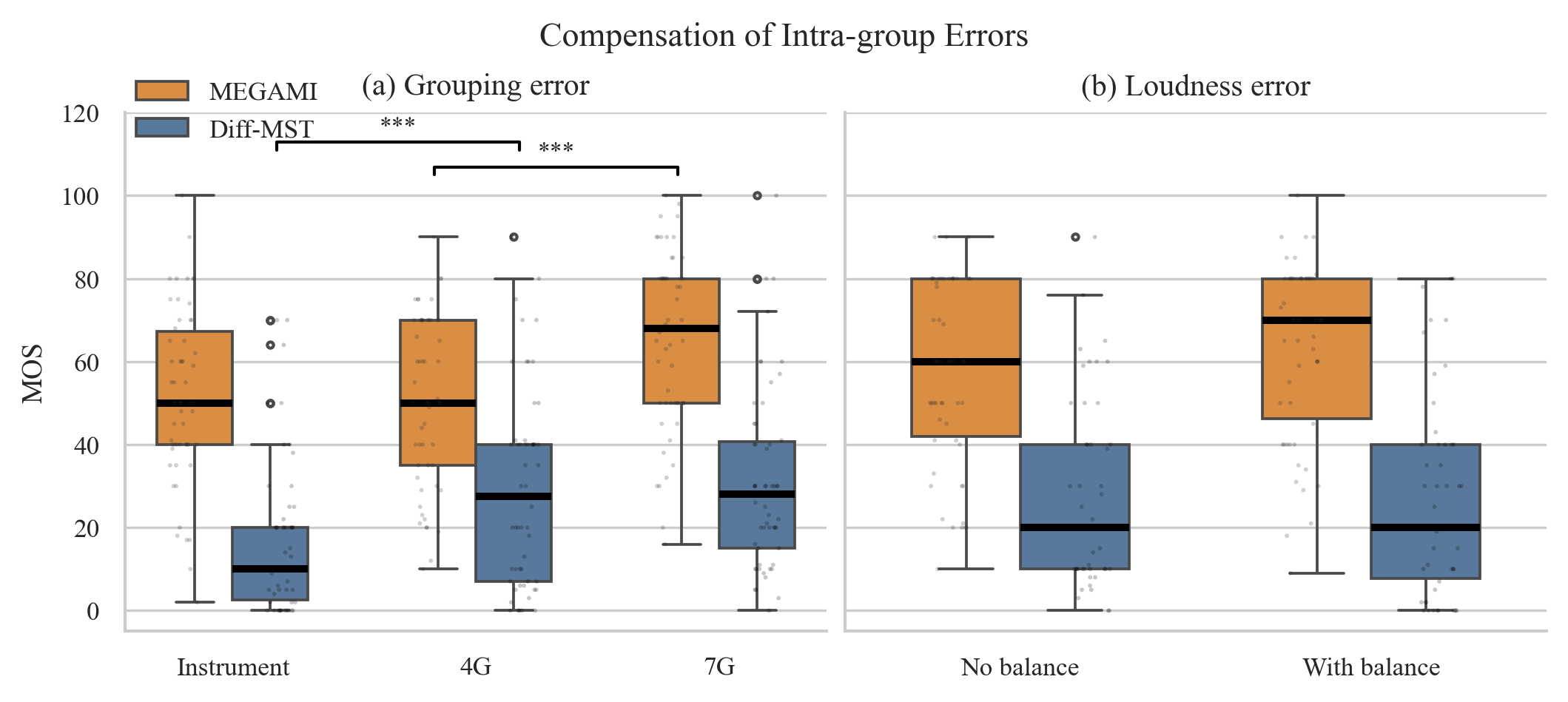}
\caption{Experiment 2. (a) Inappropriate grouping degrades downstream performance. MEGAMI benefits most from the 7-group condition, while Diff-MST mainly benefits from avoiding instrument-based grouping. (b) Loudness errors show weaker and less consistent effects.}
\label{fig:rq2}
\end{figure*}

\subsection{Experiment 2: Compensation of Intra-group Errors}

\subsubsection{Grouping errors}

As shown in Fig.~\ref{fig:rq2}(a), for MEGAMI, 7-group significantly outperforms 4-group (65.39 vs.\ 50.59, $q=6.44\times10^{-6}$, $d_z=0.70$, $n=54$). For Diff-MST, instrument-based grouping performs significantly worse than 4-group (16.81 vs.\ 29.37, $q=1.55\times10^{-4}$, $d_z=0.56$), whereas 4-group and 7-group are comparable (29.37 vs.\ 30.91, $q=0.618$, $d_z=0.08$). Inappropriate grouping therefore reduces the structural quality available to downstream models, although the preferred grouping granularity is model-dependent.

\subsubsection{Loudness errors}
For MEGAMI, the no-balance condition receives a lower mean score than the with-balance condition (57.13 vs.\ 63.39), but the difference is not significant after correction ($q=0.151$, $d_z=0.21$, $n=54$). Diff-MST shows no meaningful difference (26.70 vs.\ 25.87, $q=0.789$, $d_z=0.04$). Loudness perturbations therefore have weaker and less conclusive effects than grouping changes in the present experiment.


\begin{figure}[t]
\centering
\includegraphics[alt={Box plot comparing full-mix quality of two-stage and single-stage models},width=\linewidth]{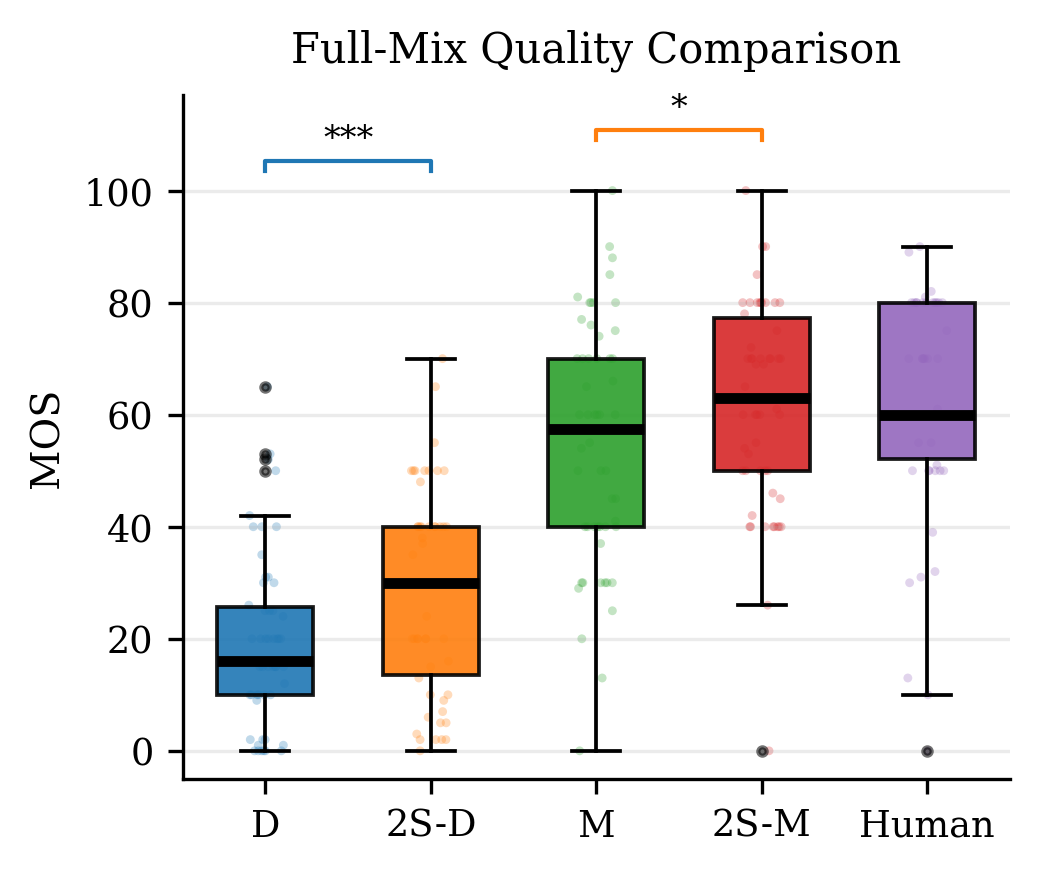}
\caption{Full-mix quality (Experiment 3). Both two-stage variants significantly outperform their corresponding single-stage baselines.}
\label{fig:rq3}
\end{figure}

\subsection{Experiment 3: Full-Mix Ablation}

Experiment 3 evaluates whether explicit two-stage decomposition improves full-mix quality (Fig.~\ref{fig:rq3}). 2S-MEGAMI significantly outperforms MEGAMI (61.57 vs.\ 54.65, $q=0.0426$, $d_z=0.29$, $n=54$). Likewise, 2S-Diff-MST significantly outperforms Diff-MST (28.69 vs.\ 19.44, $q=7.25\times10^{-4}$, $d_z=0.50$, $n=54$). These within-family comparisons provide direct evidence that explicit two-stage decomposition improves full-mix quality for both evaluated model families. Table~\ref{tab:key_results_all} summarizes the key paired comparisons discussed across Experiments~1--3.

\begin{table}[t]
\centering
\caption{Key comparisons across the three experiments. Columns A and B show the corresponding mean ratings. M = MEGAMI; D = Diff-MST; I = instrument-based grouping; 4G/7G = four-/seven-group grouping; NB/WB = no-balance/with-balance; and 2S = two-stage. Here, $q$ denotes the FDR--BH-adjusted $p$-value, $|d_z|$ denotes the absolute paired-samples Cohen's effect size, and $n$ is the number of valid paired listener--item observations.}
\label{tab:key_results_all}
\scriptsize
\setlength{\tabcolsep}{2.6pt}
\resizebox{\columnwidth}{!}{%
\begin{tabular}{llrrrrr}
\toprule
Exp. & Comp. & A & B & $q$ & $|d_z|$ & $n$ \\
\midrule
1  & ELL -- M    & 61.08 & 55.47 & 0.2623   & 0.21 & 36 \\
1  & ELL -- D    & 61.08 & 21.36 & 2.68e-08 & 1.23 & 36 \\
1  & M -- D      & 55.47 & 21.36 & 1.64e-09 & 1.41 & 36 \\
\midrule
2a& 7G-M --4G-M & 65.39 & 50.59 & 6.44e-06 & 0.70 & 54 \\
2a& I-D -- 4G-D  & 16.81 & 29.37 & 1.55e-04 & 0.56 & 54 \\
2a& 4G-D -- 7G-D & 29.37 & 30.91 & 0.6179   & 0.08 & 54 \\
\midrule
2b& NB-M--WB-M & 57.13 & 63.39 & 0.1506   & 0.21 & 54 \\
2b&NB-D --WB-D & 26.70 & 25.87 & 0.7885   & 0.04 & 54 \\
\midrule
3  & 2S-M -- M    & 61.57 & 54.65 & 0.0426   & 0.29  & 54 \\
3  & 2S-D -- D    & 28.69 & 19.44 & 7.25e-04 & 0.50  & 54 \\
\bottomrule
\end{tabular}%
}
\end{table}

\section{Discussion}

\subsection{Why Full-Mix Models Do Not Necessarily Transfer to Subtasks}

Experiment 1 shows that a structurally simpler task is not necessarily easier for a model trained for full mixing. Intra-group processing reduces the number of interactions but emphasizes different objectives, including local balance and within-group clarity. The contrast between MEGAMI and Diff-MST suggests that successful transfer depends on model formulation, training conditions, and alignment with the target subtask. The present comparison is limited to two learned model families, so the architectural causes of this difference require further study.

\subsection{Why Some Intra-group Errors Are Difficult to Recover Downstream}

Experiment 2 highlights the importance of early-stage structure. Grouping determines how tracks are aggregated before the inter-group model receives them, and inappropriate assignments can remove useful functional relationships. Loudness effects are less clear: MEGAMI shows a non-significant directional decrease under incorrect balance, whereas Diff-MST remains similarly low in both conditions. For Diff-MST, the uniformly low ratings may limit the sensitivity of the comparison, making it difficult to distinguish genuine downstream compensation from a possible floor effect. The current evidence therefore supports grouping as a core design component, while the role of intra-group loudness structure requires broader evaluation.

\subsection{Implications for Future Automatic Mixing Systems}

Experiment 3 shows that decomposition can improve existing models without requiring a newly trained hierarchical architecture. The strong intra-group performance of ELL also indicates that local processing need not rely on computationally expensive neural models. Future systems may combine efficient task-specific processing for well-defined local objectives with learned models for global coordination. The central implication is therefore a design principle: separating local balance optimization from global mix coordination provides a flexible structure for automatic mixing.

\subsection{Limitations and Scope of the Findings}

The study evaluates three densely arranged pop and rock excerpts, enabling consistent comparison across many conditions but limiting musical diversity and song-level generalization. We did not intentionally restrict participants based on their background. However, evaluating subtle mixing differences requires listeners to attend to technical aspects such as balance, clarity, and spatial relationships, which can be challenging for listeners without relevant experience. Among the recruited music enthusiasts, only a subset were able to complete the full listening test and pass the reliability assessment. The retained listeners mainly had music-production or mixing experience; their assessments are suitable for subtle technical comparisons but may not represent general-audience preferences. Finally, grouping and model stages were executed manually according to predefined rules. Although this supports reproducibility, automatic functional grouping and integration into a unified end-to-end application remain important directions for future work.

\section{Conclusion}

We present a subtask-oriented analysis of two-stage automatic mixing. The evaluated full-mix models show different transfer behavior on intra-group mixing; inappropriate grouping causes clear downstream degradation, while altered loudness relationships have weaker and model-dependent effects. Most importantly, 2S-MEGAMI and 2S-Diff-MST both significantly outperform their corresponding single-stage baselines.

These findings suggest that automatic mixing systems may benefit from explicitly separating intra-group processing from inter-group coordination and treating grouping and local balance as core design components rather than secondary preprocessing steps.

\section{Ethics Statement}

All listening tests were conducted in accordance with standard ethical guidelines for human-subject research. Participants provided informed consent, could withdraw at any time, and no personally identifiable data were retained.

\bibliography{refsnew}

%
%
%
%

\end{document}